\pdfoutput=1
\documentclass{jpp}

\usepackage{graphicx}
\usepackage{natbib}
\usepackage[T1]{fontenc}
\usepackage{bm}
\usepackage{color}
\usepackage{graphicx}
\usepackage{amsmath}
\usepackage{amssymb}
\usepackage{amsfonts}
\usepackage{courier}
\usepackage{txfonts}
\ifCUPmtlplainloaded \else
  \checkfont{eurm10}
  \iffontfound
    \IfFileExists{upmath.sty}
      {\typeout{^^JFound AMS Euler Roman fonts on the system,
                   using the 'upmath' package.^^J}%
       \usepackage{upmath}}
      {\typeout{^^JFound AMS Euler Roman fonts on the system, but you
                   dont seem to have the}%
       \typeout{'upmath' package installed. JPP.cls can take advantage
                 of these fonts, if you use 'upmath' package.^^J}%
       \providecommand\upi{\pi}%
      }
  \else
    \providecommand\upi{\pi}%
  \fi
\fi

\ifCUPmtlplainloaded \else
  \checkfont{msam10}
  \iffontfound
    \IfFileExists{amssymb.sty}
      {\typeout{^^JFound AMS Symbol fonts on the system, using the
                'amssymb' package.^^J}%
       \usepackage{amssymb}%

      }{}
  \fi
\fi

\ifCUPmtlplainloaded \else
  \IfFileExists{amsbsy.sty}
    {\typeout{^^JFound the 'amsbsy' package on the system, using it.^^J}%
     \usepackage{amsbsy}}
    {\providecommand\boldsymbol[1]{\mbox{\boldmath $##1$}}}
\fi

\newcommand\eg{e.g.\ }

\usepackage[T1]{fontenc}
\usepackage{bm}
\usepackage{color}
\usepackage{graphicx}
\newcommand{\Eq}[1]{Eq.~\eqref{#1}}

\newcommand{\sub}[1]{\ensuremath{_{\text{#1}}}}
\newcommand{\rd}{\ensuremath{\mathrm{d}}}
\newcommand{\Ordo}{\mathcal{O}}

\title[Large-angle collision operator for runaway avalanches]{On the relativistic large-angle electron collision operator for runaway avalanches in plasmas}

\author[O. Embr\'eus, A. Stahl, T. F\"ul\"op]
{O.\ns E\ls M\ls B\ls R\ls \'E\ls U\ls S,
\ns
A.\ns S\ls T\ls A\ls H\ls L \ns
\and T.\ns F\ls \"{U}\ls L\ls \"{O}\ls P}

\affiliation{Department of Physics, Chalmers University of Technology, Gothenburg, SWEDEN}

\begin{document}

\maketitle

\begin{abstract}  
  Large-angle Coulomb collisions lead to an avalanching generation of
  runaway electrons in a plasma. We present the first fully
  conservative large-angle collision operator, derived from the
  relativistic Boltzmann operator.  The relation to previous models
  for large-angle collisions is investigated, and their validity
  assessed.  We present a form of the generalized collision operator
  which is suitable for implementation in a numerical kinetic-equation
  solver, and demonstrate the effect on the runaway-electron growth
  rate. Finally we consider the reverse avalanche effect, where
  runaways are slowed down by large-angle collisions, and show that
  the choice of operator is important if the electric field is close
  to the avalanche threshold.
\end{abstract}

\maketitle

\section{Introduction}
Large-angle collisions are associated with large momentum transfers, however their influence can often be ignored in plasma physics, as
the cumulative effect of many small-angle deflections are larger by a
factor of the Coulomb logarithm,
$\ln\Lambda$~\citep{Rosenbluth1957,trubnikov}.  In many plasmas,
e.g.~magnetic fusion plasmas and astrophysical plasmas, $\ln\Lambda$
is typically of order 10-30. This allows collisions to be accurately
accounted for using a Fokker-Planck equation, originally derived for
Coulomb interactions by Landau as the small-momentum-transfer limit of
the Boltzmann equation~\citep{landauCoulomb}.

A unique situation occurs in runaway acceleration of electrons, where
large-angle collisions can play a dominant role even for large
$\ln\Lambda$, as they cause an exponential growth of the runaway
density -- a {\em runaway avalanche}~\citep{jayakumar1993}.  Runaway is the acceleration of
particles in the presence of an electric field which exceeds the
critical field $E\sub{c} = n\sub{e} \ln\Lambda e^3 (4\upi\varepsilon_0^2 m\sub{e}
c^2)^{-1}$ \citep{ConnorHastie1975}, where $n\sub{e}$ is the electron
density, $e$ is the elementary charge, $\varepsilon_0$ is the vacuum permittivity, $m\sub{e}$ is the electron rest mass, and $c$ is the speed of light. 
Since the collisional drag for superthermal electrons is
given by $F\sub{c} = e E\sub{c} (v/c)^{-2}$, any electrons with speed greater
than the critical speed $v\sub{c} = c\sqrt{E\sub{c}/E}$ will be accelerated
indefinitely, and are hence referred to as runaway
electrons~\citep{Wilson1925}.  Runaway electrons occur in a wide range
of plasmas, \eg in atmospheric discharges~\citep{dwyer}, in solar
flares~\citep{Holman1985}, and in tokamak disruptions when the plasma
current changes quickly and a strong electric field is induced
\citep{Gill1993,Jaspers1993}. Due to the large plasma current they would carry,
 reactor-scale tokamaks such as ITER will be
particularly susceptible to the conversion of plasma current to
relativistic runaway-electron current by large-angle collisions during disruptions~\citep{RosenbluthPutvinski1997}. The
subsequent uncontrolled loss of a runaway-electron beam could damage plasma-facing
components, and the runaways therefore pose a critical threat to the
viability of nuclear fusion for energy production~\citep{hollmann15iter}.

In a plasma, runaways are mainly generated by two separate mechanisms. When
the electrons in the runaway region $v > v_c$ are being accelerated,
collisional velocity-space diffusion will feed thermal electrons into
the runaway region at a steady rate.  This \emph{primary runaway
  generation}, or \emph{Dreicer} mechanism~\citep{Dreicer1960},
generates new runaways at a rate which is exponentially sensitive to
the electric field.  The runaway population growth rate was derived in
\citep{ConnorHastie1975,Cohen} and is
\begin{align*}
&\left(\frac{\rd n\sub{RE}}{\rd t}\right)\sub{prim}\hspace{-2mm} \approx \kappa
\frac{n\sub{e}}{\tau\sub{c}} \left(\frac{E}{E\sub{D}}\right)^{-\frac{3}{16}(1+Z\sub{eff})h}\hspace{-3mm}
\exp\left[
	-\lambda\frac{E\sub{D}}{4E}-\sqrt{\eta \frac{(1+Z\sub{eff})E\sub{D}}{E}}
\right], 
\end{align*}
where the undetermined constant factor $\kappa$ is of order unity. Here $n\sub{RE}$ is the number density of runaways, $\tau\sub{c} = 4\upi\varepsilon_0^2 m\sub{e}^2 c^3/n\sub{e} e^4 \ln\Lambda$ is the relativistic-electron collision time, $E\sub{D} = m\sub{e}^2 c^3/(e \tau\sub{c} T\sub{e})$ is the so-called \emph{Dreicer field}, and $Z\sub{eff} = \sum_i n_i Z_i^2 / \sum_i n_i Z_i$ is the effective ion charge (with the sum taken over all ion species $i$). The parameters $h$, $\lambda$ and $\eta$ (not given here) depend on $E/E\sub{c}$ and approach unity as $E/E\sub{c}$ becomes large (in the non-relativistic limit), but ensure that the growth rate vanishes as $E\to E\sub{c}$.

A \emph{secondary runaway generation} mechanism is provided by
large-angle collisions, whereby an electron with kinetic energy 
$\epsilon = (\gamma-1)m\sub{e} c^2> 2\epsilon\sub{c}$ 
can send a stationary target electron into the runaway
region in a single collision event while remaining a runaway itself, where $\epsilon\sub{c}$ is the kinetic energy corresponding to the critical speed.
Secondary generation, also referred to as avalanche generation due to
the resulting exponential growth of the runaway population, generates
new runaways at a rate calculated by~\citet{RosenbluthPutvinski1997}
to be approximately
\begin{align}
\left(\frac{\rd n\sub{RE}}{\rd t}\right)\sub{ava}\!\! \approx C\frac{n\sub{RE}}{2\ln\Lambda\tau\sub{c}} \left(\frac{E}{E\sub{c}} -1 \right).
\label{eq:ava_growth_rate}
\end{align}
The function $C = C(E,\,Z\sub{eff})$ was shown to be $C = 1$ when
collisional diffusion is neglected (formally by setting $Z\sub{eff} = -1$). 

While the avalanche growth rate is formally of order $1/\ln\Lambda$
smaller than the primary generation rate, the more favorable scaling
with electric field makes it the dominant source of new runaways for
sufficiently large runaway populations $n\sub{RE}$ or sufficiently
small $E/E\sub{D}$, i.e.~at sufficiently low temperature.  In the presence
of a constant electric field $E$, with no initial runaway population (apart from a small primary seed),
the secondary generation rate will exceed the primary one after
approximately one avalanche e-folding time $t\sub{ava} \approx
2\ln\Lambda m\sub{e} c/[Ce(E-E\sub{c})]$. This corresponds to the time when the
fastest runaway has been accelerated to a kinetic energy $E\sub{k}\approx
\ln\Lambda/C$ MeV~\citep{jayakumar1993} (neglecting the weak
electric-field dependence of $C$). Numerically, $t\sub{ava}
\approx 3.4 \ln\Lambda/[C(E-E\sub{c})]$\,ms, with $E$ and $E\sub{c}$ in V/m. If
the electric field decreases in magnitude with time, avalanche will
become important even earlier.  In many practical runaway scenarios,
the runaway process will last for multiple
$t\sub{ava}$~\citep{Gurevich1994,GurevichZybin2001,Helander2002},
and secondary generation will therefore be the dominant runaway
mechanism.

In this work we derive a conservative large-angle (also known as
``close'' or ``knock-on'') collision model from the high-energy limit
of the linearized relativistic Boltzmann collision integral. We will
show how the operators used to model large-angle
collisions in previous studies are obtained through various approximations 
of the Boltzmann collision operator, and how our more general operator resolves 
issues with previous models and allows the study of new
physical effects. In particular, we resolve the issue
of double-counting large-angle and small-angle collisions, and show
that this development is essential to accurately capture the dynamics. We find
that the change to the runaway growth rate due to the new operator is
largest during the early stages of the runaway acceleration process,
and the likelihood of a given runaway seed transforming into a serious
runaway beam can thus potentially be affected.  Furthermore, we 
consider the effect of the inverse knock-on process, where a runaway
is slowed down in a single large-angle collision. This effect was
recently shown by~\citet{Aleynikov2015} to be significant for runaway 
in a near-threshold electric field.

The rest of the paper is organized as follows. In Section 2 we
introduce the theoretical models describing the large-angle collisions. After
giving an overview of the existing models, we present a
derivation of the new conservative operator. In Section 3 we 
investigate the effect of the new operator on the runaway growth
rate numerically, using the kinetic-equation solver \textsc{CODE}
~\citep{CODEpaper2014,Stahl2016}.  Finally,
we summarize our conclusions in Section 4.

\section{Theoretical models for runaway generation due to large-angle collisions}
One of the earliest models for avalanche runaway generation was
introduced by \citet{RosenbluthPutvinski1997}. Due to its simple form, 
suitable for analytical development, it has been widely used to study 
the dynamics of an avalanching runaway population~\citep{ARENA,Smith2005,Fulop2006,Nilsson2015}.  Rosenbluth and Putvinski proposed a kinetic
equation for the electron distribution of the form
\begin{align}
\frac{\rd f\sub{e}}{\rd t} = C\sub{FP}(f\sub{e}) + S(f\sub{e}),
\label{eq:colop split}
\end{align}
where $\rd f\sub{e}/\rd t$ represents the advective part of the motion,
$C\sub{FP}$ is the Fokker-Planck collision operator and $S$ a source
term representing ``\emph{secondary high energy electrons knocked out
  of their orbits by close collisions of a primary relativistic
  electron with low energy electrons from the background
  plasma}''~\citep{RosenbluthPutvinski1997}. Assuming all existing
runaways to be infinitely energetic and having zero pitch angle, they
obtained (here adapting their more general result to a homogeneous plasma)
\begin{align}
S\sub{RP}(p,\,\xi,\,\varphi) = \frac{n\sub{RE}}{4\upi \tau\sub{c} \ln\Lambda} \delta(\xi - \xi_0) \frac{m\sub{e}^3c^3}{p^2}\frac{\partial}{\partial p}\left(\frac{1}{1-\gamma}\right),
\label{eq:RP source}
\end{align}
where $\xi = \cos\theta = p_\parallel / p$ is the pitch-angle cosine,
$\gamma = \sqrt{1+(p/m\sub{e} c)^2}$ is the Lorentz factor,
$\xi_0 = \sqrt{(\gamma-1)/(\gamma+1)}$ and the momentum-space volume
element is $p^2 \rd p \rd \xi \rd \varphi$, with $\varphi$ the
azimuthal angle of the momentum (the gyroangle). The delta function
ensures that secondary electrons are only born on the parabola
$p_\perp^2 = 2p_\parallel m\sub{e} c$ in momentum space. In the
non-relativistic limit, $p \ll m\sub{e} c$, secondaries are born at
perpendicular angles, $p_\parallel \sim 0$, and are prone to trapping
in an inhomogeneous magnetic field. Away from the magnetic axis of a
tokamak, this can lead to a strong reduction in the avalanche growth
rate, as recently shown by~\citet{Nilsson2015}.

A more general model was later described by~\citet{Chiu1998} (from now
on referred to as the Chiu-Harvey operator), which has also been used
in runaway studies~\citep{Chiu1998,CQL2,Stahl2016}. Allowing runaway-electron
energies to be finite but assuming the runaway pitch-angle to be zero, they
obtained a knock-on source term
\begin{align}
S\sub{CH}(p,\,\xi,\,\varphi) &= \frac{1}{4\upi \tau\sub{c} \ln\Lambda } \frac{p_1^2}{m\sub{e} c p\gamma \xi}F(p_1,\,t)\Sigma(\gamma,\,\gamma_1'), 
\label{eq:CH source}  \\
\Sigma(\gamma,\,\gamma_1) &= \frac{\gamma_1^2}{(\gamma_1^2-1)(\gamma-1)^2(\gamma_1-\gamma)^2}
\Biggr( (\gamma_1-1)^2 - \frac{(\gamma-1)(\gamma_1-\gamma)}{\gamma_1^2} \nonumber \\
&\hspace{7mm} \times 
\Big[2\gamma_1^2+2\gamma_1-1-(\gamma-1)(\gamma_1-\gamma)\Big]\Biggr), 
\label{eq:moller}
\end{align}
where $\Sigma = (2\upi r_0^2)^{-1} \rd \sigma/\rd \gamma$ is the
normalized M\o{}ller differential cross-section for free-free
electron-electron scattering \citep{Moller1932}, $r_0 = e^2/(4\upi\varepsilon_0 m\sub{e} c^2)$ is
the classical electron radius, and $\gamma_1$ is connected to $p$ and $\xi$ by the
relation
\begin{align}
\xi \equiv \xi^*(\gamma,\,\gamma_1) &= \sqrt{\frac{\gamma_1+1}{\gamma_1-1}\frac{\gamma-1}{\gamma+1}} \Leftrightarrow p_1 = \frac{2p\xi^*}{1+\xi^{*2} - \gamma(1-\xi^{*2})} ,
\label{eq:chiu harvey kinematics}
\end{align}
where a misprint in the original paper incorrectly replaced the
$\gamma-1$ factor with $\gamma_1$. Since the authors work under the
assumption that the runaway pitch-angles are negligible, the
distribution only appears in the angle-averaged form
\begin{align}
F(p_1,\,t) = \int \rd \xi_1 \rd \varphi_1  \, p_1^2 f\sub{e}(p_1,\,\xi_1,\,t).
\end{align}

Both models for large-angle collisions presented above
suffer from several defects.  In particular, they do not
conserve particle number, energy or momentum.  
In addition, the Rosenbluth-Putvinski model assumes that the incoming particle momentum
is infinite, which has the consequence that particles can be created
with an energy higher than any of the existing runaways.  This assumption is not made in the model derived by \citet{Chiu1998}, where the electron energy distribution is properly taken into account, but all incident runaways are still assumed to have
zero pitch-angle. The magnitudes of both
sources ($S\sub{RP}$ and $S\sub{CH}$) increase rapidly with decreasing momenta and the sources 
are thus sensitive to the choice of cut-off momentum (introduced to avoid double-counting small-angle collisions). 

In the following we will derive a knock-on collision model from the
Boltzmann collision integral. As we will show, the model takes into
account the full momentum dependence of the primary distribution, and
conserves particle number, momentum and energy, while also consistently distinguishing between small and large-angle collisions, therefore avoiding double-counting.

\subsection{The Boltzmann collision integral and the Fokker-Planck limit}
The Boltzmann collision operator gives the time-rate-of-change of the
distribution function due to binary collisions, described by an
arbitrary differential cross-section. 
It can be derived with the following heuristic
argument~\citep{CercignaniKremer,Montgomery}.
The collision operator can be defined as $C_{ab}(f_a) = (\rd
n_a)_{\mathrm{c},ab}/\rd t \rd p$, where $(\rd n_a)_{\mathrm{c},ab}$ is the differential
change in the density of a species $a$ due to collisions with species
$b$, and is defined in terms of the differential cross-section $\rd
\sigma$ by~\citep{CercignaniKremer,LLKinetics}
\begin{align}
(\rd n_a)_{\mathrm{c},ab} = f_a(\boldsymbol{p}_1)f_b(\boldsymbol{p}_2) \bar{g}_{\text\o} \rd \bar\sigma_{ab} \rd\boldsymbol{p}_1\rd\boldsymbol{p}_2 \rd t 
- f_a(\boldsymbol{p})f_b(\boldsymbol{p}') g_{\text\o} \rd \sigma_{ab} \rd \boldsymbol{p}\,\rd\boldsymbol{p}'\rd t.
\label{eq:diff n}
\end{align}
The first term on the right-hand side, the gain term, describes the rate at which particles
$a$ of momentum $\boldsymbol{p}_1$ will scatter to momentum $\boldsymbol{p}$.  The
second term, the loss term, is the rate at which particles $a$ scatter
away from momentum $\boldsymbol{p}$.  Here, we introduced the M\o{}ller
relative speed $g_\text\o = \sqrt{(\boldsymbol{v}-\boldsymbol{v}')^2 - (\boldsymbol{v}\times
  \boldsymbol{v}')^2/c^2}$ and the differential cross-section $\rd\sigma_{ab}$
for scattering events $\boldsymbol{p},~\boldsymbol{p}' \to \boldsymbol{p}_1,~\boldsymbol{p}_2$. The
barred quantities are defined likewise, but with $\boldsymbol{p}$ exchanged 
for $\boldsymbol{p}_1$ and $\boldsymbol{p}'$ for $\boldsymbol{p}_2$.
Since the
interactions are viewed as instantaneous, the time labels of the
distribution functions have been suppressed for clarity of notation.

The elastic differential cross-section satisfies the symmetry
property 
$\bar{g}_{\text\o}\rd \bar\sigma_{ab} \rd\boldsymbol{p}_1\rd\boldsymbol{p}_2 =
g_{\text\o}\rd\sigma_{ab} \rd\boldsymbol{p}\rd\boldsymbol{p}'$ (known as the \emph{principle of detailed balance}~\citep{weinberg_fields}), allowing the collision operator
to be cast in the commonly adopted symmetric form
\begin{align}
C_{ab}^{B} = \int \rd \boldsymbol{p}' \rd \sigma_{ab} \, g_{\text\o}\Big[f_a(\boldsymbol{p}_1)f_b(\boldsymbol{p}_2) - f_a(\boldsymbol{p})f_b(\boldsymbol{p}')\Big],
\label{eq:classic boltz}
\end{align}
where $\boldsymbol{p}_1$ and $\boldsymbol{p}_2$ (six degrees of freedom) are uniquely
determined in terms of $\boldsymbol{p}$ and $\boldsymbol{p}'$ by two scattering angles
and four constraints by the conservation of momentum and energy,
\begin{align}
\boldsymbol{p}_1 + \boldsymbol{p}_2 &= \boldsymbol{p} + \boldsymbol{p}', \\
m_a \gamma_1 + m_b \gamma_2 &= m_a \gamma + m_b \gamma'.
\end{align}
From this collision operator, the Fokker-Planck operator, which is
often used in plasma physics, can be obtained by a Taylor expansion to
second order in the momentum transfer $\Delta\boldsymbol{p} = \boldsymbol{p}_1 -
\boldsymbol{p}$~\citep{landau1936,akama}, motivated by the fact that the
cross-section for Coulomb collisions is singular for small
deflections. It is then seen that the contribution of small-angle
collisions is larger than those of large-angle collisions by a factor
of the Coulomb logarithm,
\begin{align}
\ln \Lambda = \int_{\cot(\theta\sub{max}/2)}^{\cot(\theta\sub{min}/2)} \frac{\rd \lambda}{\lambda} = \ln \left(\cot\frac{\theta\sub{min}}{2} \right),
\label{eq:lnLambda}
\end{align}
where the maximum center-of-mass deflection angle for self collisions
is $\theta\sub{max} = \upi/2$ (not $\upi$ as it is for unlike-species
collisions, or collisions would be double counted) and
$\theta\sub{min}$ is a cut-off required to regularize the expression,
typically chosen as the scattering angle corresponding to impact
parameters of order the Debye length\footnote{{In the quantum-mechanical treatment, it is rather the de-Broglie wavelength of the center-of-mass momentum transfer $\lambda = \hbar/|\boldsymbol{p}^*_1-\boldsymbol{p}^*|$ that cuts off at the Debye length.}}, beyond which particles will not
interact because of Debye screening.

Note that by using a total collision operator $C\sub{FP} + C\sub{Boltz}$ as prescribed by \Eq{eq:colop split}, the
Boltzmann operator has effectively been added twice, although different approximations are used to evaluate the two terms. A subset of collisions will therefore be double counted.
One way to resolve this issue
is to apply the Fokker-Planck operator only to collision angles
smaller than some $\theta\sub{m}$, and the knock-on (Boltzmann) operator for $\theta > \theta\sub{m}$.  The Coulomb logarithm used in the Fokker-Planck operator then
ought to be changed from Eq.~(\ref{eq:lnLambda}) to
\begin{align}
\overline{\ln\Lambda} = \ln\Lambda - \ln\left(\cot\frac{\theta\sub{m}}{2}\right).
\label{eq:lnLambda mod}
\end{align}
When an incident electron of momentum $p$ knocks a stationary electron
to momentum $p\sub{m}$, the corresponding center-of-mass scattering angle
$\theta\sub{m}$ is given by
\begin{align}
\cot\frac{\theta\sub{m}}{2} &= \sqrt{\frac{\gamma-\gamma\sub{m}}{\gamma\sub{m}-1}}.
\end{align}
By using this energy-dependent modification to the Coulomb logarithm,
no collisions will be double counted.  Indeed, by taking the energy
moment of the test-particle collision operator (the sum of
Fokker-Planck and Boltzmann), it can be verified that with this
choice, the average energy-loss rate experienced by a test particle
becomes independent of the cut-off $p\sub{m}$, when
$p\sub{m} \ll m\sub{e} c$.

The number of collisions that are double-counted can often be
significant when this effect is unaccounted for. Assuming $v\sub{m}/c \sim
\sqrt{E\sub{c}/E}$ to be located at a non-relativistic energy (that is, we
assume $E\gg E\sub{c}$), the modification to the Coulomb logarithm is
approximately given by $\ln \hspace{-1.3mm}\sqrt{ 2(E/E_c)(\gamma-1)}$. For
highly energetic electrons with $\gamma\sim 50$ and $E/E\sub{c} \sim 100$,
this corresponds to a change of approximately 5, which -- depending on
plasma parameters -- typically constitutes a relative change to the
Coulomb logarithm of $25$-$50\%$.

In principle, as $\theta\sub{m}$ approaches the cut-off imposed by Debye
screening (or the binding energy of atoms in the case of electrons in
neutral media), the Boltzmann operator will account for all collisions
and $\overline{\ln\Lambda} = 0$.  However, this corresponds to a
cut-off momentum smaller than thermal, $p\sub{m} \ll p\sub{Te}$, and the
assumption of stationary targets is violated when evaluating the
operator in the bulk region.  In addition, to numerically resolve the
Boltzmann operator in a finite-difference scheme, the grid spacing in
momentum must be much smaller than $p\sub{m}$, and it is therefore
desirable to choose $p\sub{m}$ as large as allowed while having a well
converged description of the secondary-generation rate. The
sensitivity of the result to the choice of $p\sub{m}$ is investigated in the
next section.

In the following we will find it more useful not to work with the
symmetric form of the Boltzmann operator given by Eq.~(\ref{eq:classic
  boltz}), but instead use the alternative given directly from
Eq.~(\ref{eq:diff n}),
\begin{align}
C_{ab}\{f_a,\,f_b\}(\boldsymbol{p}) =& \int \rd \boldsymbol{p}_1 \int \rd \boldsymbol{p}_2 \, \frac{\partial \bar\sigma_{ab}}{\partial \boldsymbol{p}} \bar g_{\text\o} f_a(\boldsymbol{p}_1)f_b(\boldsymbol{p}_2) \nonumber \\
&- f_a(\boldsymbol{p}) \int \rd \boldsymbol{p}' \, g_{\text\o}f_b(\boldsymbol{p}') \sigma_{ab}(\boldsymbol{p},~\boldsymbol{p}') ,
\end{align}
where $\sigma_{ab}(\boldsymbol{p},~\boldsymbol{p}') = \int \rd \boldsymbol{p}_1 \, \partial \sigma_{ab}/\partial \boldsymbol{p}_1$ is the total cross-section.

\subsection{Derivation of a conservative knock-on operator}
For the avalanche problem, one is concerned with the electron-electron
Boltzmann operator.  We consider the scenario where a small
runaway population has been accelerated by an electric field (or other
mechanism), leaving a largely intact thermal bulk population.  We may
then write our electron distribution as $f\sub{e}(\boldsymbol{p}) = f\sub{Me}(\boldsymbol{p}) +
\delta f\sub{e}(\boldsymbol{p})$, where the runaway distribution $\delta f\sub{e}$ is
much smaller than the bulk distribution $f\sub{Me}$, $||\delta f\sub{e}|| \ll
||f\sub{Me}||$ (for example in terms of number densities $n\sub{RE} \ll
n\sub{e}$).  We may then linearize the
bilinear Boltzmann operator by ignoring terms quadratic in $\delta
f\sub{e}$, obtaining
\begin{align}
C\sub{ee}^{B}\{f\sub{e},\,f\sub{e}\} \approx C\sub{ee}^{B}\{f\sub{e},\,f\sub{Me}\} + C\sub{ee}^{B}\{f\sub{Me},\,f\sub{e}\} \equiv  C\sub{boltz}(\boldsymbol{p}),
\end{align}
where terms $C\sub{ee}\{f\sub{Me},\,f\sub{Me}\}$ vanish since $f\sub{Me}$ is
chosen as an equilibrium distribution.  The first term, the
test-particle term, describes the effect of large-angle collisions on
the runaway electrons as they collide with the thermal particles. The
second term, the field-particle term, describes the reaction of the
bulk as they are being struck by the runaways. Intuitively, one could
expect this field-particle term to constitute the avalanche knock-on
source. We shall show below that this is indeed the case.

Before giving the explicit forms of the collision operator, we will
make one final approximation. We assume that both the incident and
outgoing electrons in the large-angle collisions are significantly
faster than the thermal speed $v\sub{Te} = \sqrt{2T\sub{e}/m\sub{e}}$, so that we may
approximate the bulk population with a Dirac delta function:
$f\sub{Me}(\boldsymbol{p}) \approx n\sub{e} \delta(\boldsymbol{p})$.  The collision operator
then takes the form
\begin{align}
C\sub{ee}^{B}\{f\sub{e},\,f\sub{Me}\} &= n\sub{e}\int_{q^* > p_1 > q_0}\!\!\! \rd \boldsymbol{p}_1 \, v_1\frac{\partial \bar\sigma\sub{ee}}{\partial\boldsymbol{p}} f\sub{e}(\boldsymbol{p}_1) 
\ -\ n\sub{e} v \sigma\sub{ee}(\boldsymbol{p}) f\sub{e}(\boldsymbol{p}), \label{eq:test-particle non-L} \\
C\sub{ee}^{B}\{f\sub{Me},\,f\sub{e}\} &= n\sub{e}\int_{p_1 > q^*}\!\!\! \rd \boldsymbol{p}_1 \, v_1\frac{\partial \bar\sigma_{ee}}{\partial\boldsymbol{p}} f\sub{e}(\boldsymbol{p}_1) 
\ -\ n\sub{e} \delta(\boldsymbol{p}) \int_{p_1>q_0} \!\!\! \rd \boldsymbol{p}_1 v_1 \sigma\sub{ee}(\boldsymbol{p}_1) f\sub{e}(\boldsymbol{p})
\label{eq:field-particle non-L} 
\end{align}
The total cross-section $\sigma\sub{ee}(\boldsymbol{p})$ is given in
Eq.~(\ref{eq:total sigma}) in  Appendix~\ref{ap:conservation}. The limiting momenta $q^*$ and $q_0$ are determined from constraints
imposed by conservation laws.  For the gain term, i.e.~the first term
in each equation, energy conservation in each collision reads $
\gamma_1 = \gamma + \gamma' - 1$, where $\gamma$ and $\gamma'$ are the
Lorentz factors of the two electrons after the collision.  The conditions
$\gamma' > \gamma$ or $\gamma' < \gamma$ determines whether $\gamma$
refers to the bulk particle or runaway particle after the collision,
respectively (note that the electrons are in fact indistinguishable,
but an artificial distinction like this must be done in order to avoid
double counting).  We therefore obtain $q^*$ from setting $\gamma' =
\gamma$ in the conservation law, giving $\gamma^* = 2\gamma-1$, which
corresponds to $q^* = m\sub{e} c\sqrt{\gamma^{*2}-1}$.

Similarly, we cannot account for all collisions, since we have assumed
the bulk particles to be much slower than the outgoing particles.  We
therefore choose to account only for those collisions where incident
and outgoing particles have momenta larger than some $p=p\sub{m} \gg
p\sub{Te}$.  Setting $\gamma' = \gamma\sub{m}$ then yields the lower limit
$\gamma_0 = \gamma + \gamma\sub{m} -1$, corresponding to $q_0 = m\sub{e} c
\sqrt{\gamma_0^2-1}$. Note that for the total operator $C\sub{Boltz}$,
the two gain terms in Eqs.~\eqref{eq:test-particle non-L} and \eqref{eq:field-particle non-L} combine into one integral, taken over all momenta
$p_1 > q_0$. The full expression is thus independent of the parameter $q^*$ which distinguishes the two outgoing particles (as is expected, since the distinction is not physically relevant for scattering of identical particles).

We can now derive explicit expressions for the collision
operator. Since there are only two degrees of freedom in the
scattering process (for example two independent scattering angles),
the differential cross-section $\partial \bar\sigma_{ee}/\partial
\boldsymbol{p}$ will invariably contain a delta function. In M\o{}ller
scattering (relativistic electron-electron scattering), the
cross-section is azimuthally symmetric (assuming the electrons to be
spin-unpolarized) and takes the form
\begin{align}
\frac{\partial \bar\sigma\sub{ee}}{\partial \boldsymbol{p}} &= \frac{r_0^2}{ (m\sub{e} c)^2 p\gamma} \delta(\cos\theta\sub{s} - \xi^*)\Sigma(\gamma,\,\gamma_1), \\
\cos\theta\sub{s} &= \frac{\boldsymbol{p}_1\cdot\boldsymbol{p}}{p_1p} \equiv \xi_s, 
\end{align}
where $\theta\sub{s}$ is the deflection angle, $\Sigma$ is defined in Eq.~(\ref{eq:moller}),  and $\xi^*$ is defined in Eq.~\ref{eq:chiu harvey kinematics}. The delta function enforces the relation between
scattering angle and energy transfer that follows from the
conservation of 4-momentum. 
The gain term then takes the form
\begin{align}
 &n\sub{e}\int \rd \boldsymbol{p}_1 \, v_1\frac{\partial \bar\sigma_{ee}}{\partial\boldsymbol{p}} f\sub{e}(\boldsymbol{p}_1) = \frac{1}{4\upi \tau\sub{c}\ln\Lambda}\frac{1}{p\gamma}\int \rd p_1 \, \frac{p_1^3}{\gamma_1}\Sigma(\gamma,\,\gamma_1)
\int \rd \xi_1\rd \varphi_1 \, \delta(\xi_s-\xi^*)f\sub{e}(\boldsymbol{p}_1).
 \label{eq:generalized S}
\end{align}
This expression (when choosing integration limits appropriate for the
field-particle term) is the generalized ``knock-on source term'' $S$,
which reduces to the expressions given by
\citet{RosenbluthPutvinski1997} and \citet{Chiu1998} -- Eqs.~\eqref{eq:RP source} and \eqref{eq:CH source}, respectively -- using appropriate
approximations, as shown in Appendix~\ref{ap:rpch}.  This connection has not
been acknowledged in previous studies, to the degree that
\citet{Chiu1998}~incorrectly ascribe the discrepancy between their
result and that of~\citet{BesedinPankratov1986} by the fact that
\emph{``The present expressions are simply a statement of the total
  rate at which electrons in different velocity space elements of
  primary electrons knock a collection of cold bulk electrons into
  velocity space elements of the secondary electrons. The expression
  in~\citep{BesedinPankratov1986} uses a Boltzmann-like integral
  operator.''} In fact, as we show in Appendix~\ref{ap:rpch}, the approaches are completely equivalent, and the discrepancy is the result of an error in the
calculation of~\citet{BesedinPankratov1986}.

There are multiple ways of carrying out the integration over the delta
function; if we assume a distribution function independent of
gyro-angle, $f\sub{e}(\boldsymbol{p}_1) = f\sub{e}(p_1,\,\cos\theta_1)$, a few convenient
expressions are given by
\begin{align}
\int_{-1}^1 \rd \xi_1 \int_0^{2\upi} \rd \varphi_1  \, \delta(\xi_s - \xi^*) f\sub{e}(p_1,\,\xi_1) 
&= \int_0^{2\upi} \rd \varphi\sub{s} \, f\sub{e}(p_1,\,\xi_1) \nonumber \\
&=2 \int_{\cos(\theta+\theta^*)}^{\cos(\theta-\theta^*)} \rd \xi_1 \,\frac{f\sub{e}(p_1,\,\xi_1)}{\sqrt{1 - \xi^{*2} - \xi_1^2 - \xi^2 + 2\xi^* \xi_1\xi }} \nonumber \\
&= 2\upi \sum_L f_L(p_1)P_L(\xi)P_L(\xi^*),
\end{align}
where we have introduced the quantities
\begin{align}
\cos\varphi\sub{s} &= \frac{\xi_1-\xi^*\xi}{\sqrt{1-\xi^{*2}}\sqrt{1-\xi^2}}, \nonumber \\
f_L(p) &= \frac{2L+1}{2}\int_{-1}^1 \rd \xi \, f\sub{e}(p,\,\xi)P_L(\xi). \nonumber
\end{align}
In particular the form involving Legendre polynomials $P_L$ is a
powerful result, as it demonstrates that the linearized Boltzmann
operator is diagonal in $L$, in the sense that if $C\sub{Boltz}(\boldsymbol{p})
= \sum_L C_L(p) P_L(\cos\theta)$, then $C_L$ depends only on $f_L$
(and not other $f_l$ with $l\neq L$). This behavior exhibits the
spherical symmetry inherent in scattering on stationary targets.
Utilizing this property leads to significant practical gains
in terms of numerical computation times. Analogous expressions in
terms of Legendre polynomials and the integration over $\varphi\sub{s}$ were
also found by~\citet{GurevichZybin2001} for the so-called ionization
integral in neutral gases.  The form of the integral taken over
$\xi_1$ was obtained by~\citet{Helander1993} in the analogous
problem of elastic nucleon-nucleon scattering, and an equivalent
formulation was also recently given by~\citet{Boozer2015}. 

The Legendre modes of the collision operator are explicitly given by
\begin{align}
C_L\{f\sub{e},\,f\sub{Me}\} &= \frac{(m\sub{e} c)^{-3}}{2\tau\sub{c}  \ln\Lambda } \frac{1}{\gamma p} \int_{q_0}^{q^*} \rd p_1 \, \frac{p_1^3}{\gamma_1} f_L(p_1)P_L(\xi^*) \Sigma(\gamma,\,\gamma_1) \nonumber \\
&\hspace{3mm} - \frac{1}{4\tau\sub{c}\ln\Lambda} \frac{v}{c} f_L(p) \int_{\gamma\sub{m}}^{\gamma + 1 -\gamma\sub{m}} \rd \gamma_1 \,\Sigma(\gamma_1,\,\gamma), 
\label{eq:full test-particle}\\
C_L\{f\sub{Me},\,f\sub{e}\} &= \frac{(m\sub{e} c)^{-3}}{2\tau\sub{c}\ln\Lambda } \frac{1}{\gamma p} \int_{q^*}^{\infty} \rd p_1 \, \frac{p_1^3}{\gamma_1} f_L(p_1)P_L(\xi^*) \Sigma(\gamma,\,\gamma_1) \nonumber \\
&\hspace{3mm} - \frac{(m\sub{e} c)^{-1}}{4\tau\sub{c}\ln\Lambda} \delta_{L,0}\frac{\delta(p)}{p^2} \int_{q_0(p\sub{m})}^\infty \rd p' \,\frac{p'^3}{\gamma'} f_0(p')  \int_{\gamma\sub{m}}^{\gamma' + 1 -\gamma\sub{m}} \rd \gamma_1 \,\Sigma(\gamma_1,\,\gamma').
\label{eq:full field-particle}
\end{align}
Note further that since we only consider those collisions where both
the incident and outgoing particles have momenta $p> p\sub{m}$, the gain
terms must only be applied for $\gamma>\gamma\sub{m}$, while the
test-particle loss term is applied for $\gamma > 2\gamma\sub{m}-1$.  In
Appendix~\ref{ap:conservation} it is explicitly demonstrated that this collision operator
conserves density, momentum and energy.
\begin{figure}
\begin{center}
\includegraphics[width=1\textwidth,trim=20mm 0 15mm 0]{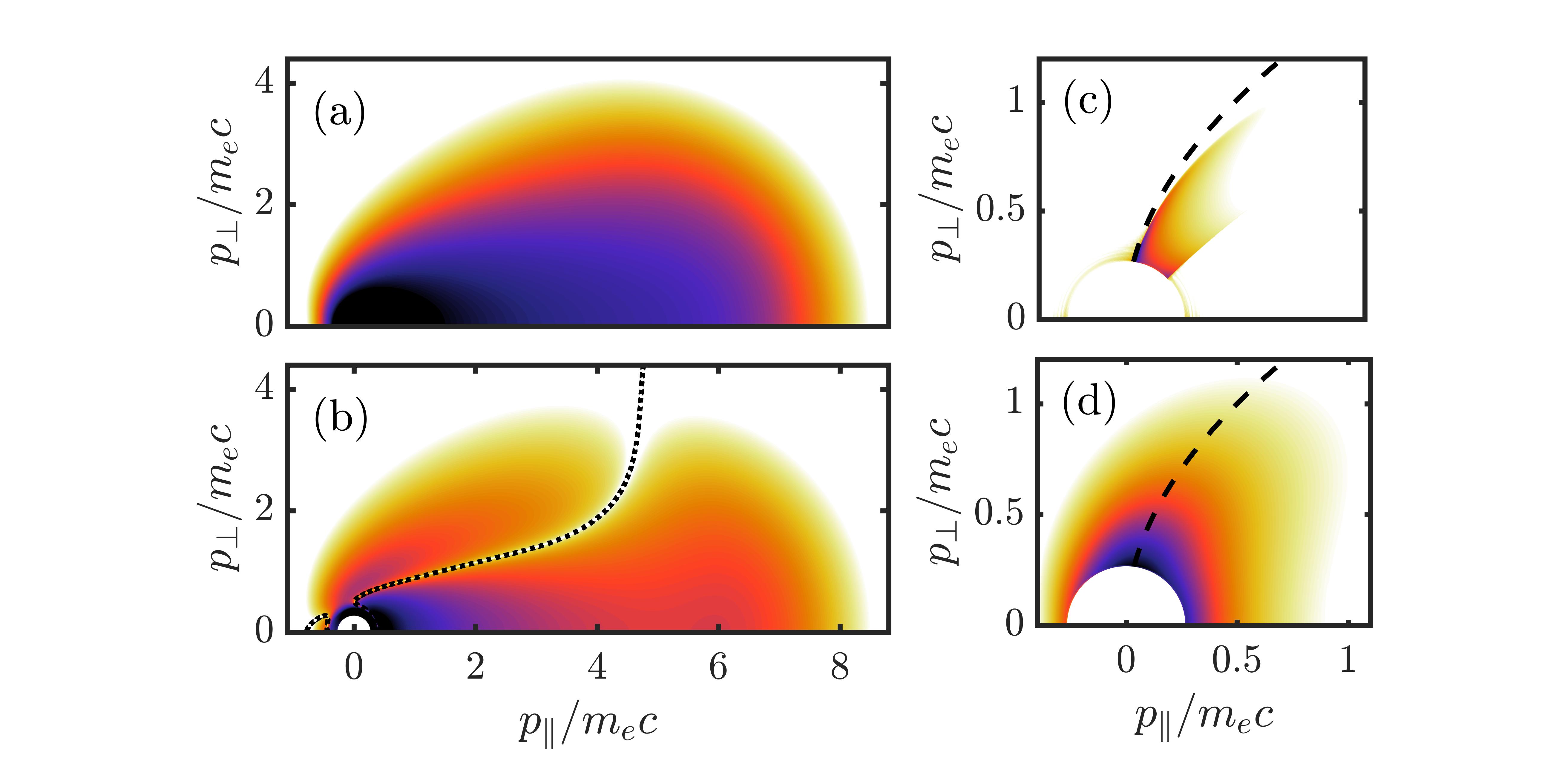}
\caption{\label{fig:knockon operators} Illustration of the large-angle
  collision operators investigated in this study. Darker colors represent larger amplitudes (in arb.~units), where white and black are separated by 3 orders of magnitude. (a) The distribution
  function $\log f\sub{e}$ with which we evaluate the large-angle collision
  operator; (c) the Chiu-Harvey operator $\log C\sub{CH}$;
  (d) the full field-particle operator $\log
  C\sub{boltz}^{\text{(fp)}}$ (dashed: the line $\xi = \sqrt{(\gamma-1)/(\gamma+1)}$, where the Rosenbluth-Putvinski
  operator creates knock-ons); (b) the magnitude of the full
  test-particle operator $\log |C\sub{boltz}^{\text{(tp)}}|$ , where
  the dotted line separates the region of negative contributions (to
  the right) from the positive contributions (to the
  left). 
   }
\end{center}
\end{figure}

A qualitative illustration of the large-angle collision operators discussed here is shown in Figure~\ref{fig:knockon operators}. A test
runaway distribution (Fig.~\ref{fig:knockon operators}a) was generated
by applying a constant electric field $E=15E\sub{c}$ for a short time
$t\approx 0.5\tau\sub{c}$ with $Z\sub{eff}=5$, and the large-angle
collision operators were evaluated in the final time step. The figures show
a snapshot of where large-angle collisions between runaways and bulk
particles create or remove electrons in phase space; comparing
Fig.~\ref{fig:knockon operators}c and Fig.~\ref{fig:knockon operators}d shows that the Chiu-Harvey operator creates secondary runaways in
a significantly smaller region in momentum space than the full
field-particle operator, however the total number of secondary runaways created is
equal between the models. Figure~\ref{fig:knockon operators}b shows the
Boltzmann test-particle operator, illustrating the reaction of the
already-present runaways: they are removed at small pitch-angles where
the runaway distribution is largest, and placed at larger pitch-angles
and lower energy. The sum of Fig.~\ref{fig:knockon operators}b and d
is the full Boltzmann operator which conserves particle number,
momentum and energy. 

Note finally that all of the knock-on models described in this paper share
the assumption of a stationary bulk, which means that the operators can only
be evaluated at speeds much larger than the thermal speed $v\sub{Te}$.  Since
the sources must be applied for speeds smaller than the critical speed
$v_c$ in order to accurately capture the runaway rate, the condition
$v\sub{Te}\ll v_c$ limits the electric-field values to $\sqrt{E} \ll
\sqrt{E\sub{D}/2}$ (effectively forming a lower limit in density and an
upper limit in temperature for a given $E$).  This limitation would be
resolved by accounting for the velocity distribution of the target
population in Eq.~(\ref{eq:diff n}), resulting in a significantly more
complicated operator.  However, in most scenarios this is not an
issue; the critical velocity tends to be significantly larger than
thermal.  If this was not the case, the entire electron population
would run away within a few collision times by the primary generation 
mechanism regardless.

\section{Numerical study of the effect of large-angle collisions}
We use the kinetic-equation solver
\textsc{CODE}~\citep{CODEpaper2014,Stahl2016} to compare the various
models for the knock-on collision operator. We use \textsc{CODE} to solve the relativistic 0D+2P kinetic equation for the electron distribution
\begin{align}
\frac{\partial f_e}{\partial t} + \left\langle\frac{\partial}{\partial \boldsymbol{p}}\cdot \bigr[ (\boldsymbol{F}\sub{L} + \boldsymbol{F}\sub{S} ) f_e \bigr]\right\rangle
= C_{ei} + C_{ee}+C_{\mathrm{boltz}}.
\label{eq:kinetic equation}
\end{align}
where $\boldsymbol{F}\sub{L}$ is the Lorentz force, and $\boldsymbol{F}\sub{S}$ is the
radiation reaction force associated with synchrotron radiation and the
brackets denote averaging over the azimuthal (gyro) angle.  $C_{ei}$
and $C_{ee}$ are the gyroaveraged Fokker-Planck collision operators
for electron-ion and electron-electron collisions, respectively.
First, we will study the sensitivity of the avalanche dynamics to the
arbitrary cut-off parameter $p\sub{m}$ and investigate the effects of
adding the test-particle Boltzmann operator, which restores
conservation laws in the knock-on collisions. We then focus on two
scenarios: (i) we revisit the classical calculation of the
steady-state avalanche growth rate in a constant electric field, (ii)
we calculate the runaway growth rate in the near-critical field,
accounting for synchrotron energy loss.

\subsection{Sensitivity to the cut-off parameter $p\sub{m}$}
\begin{figure}
\begin{center}
\includegraphics[width=0.6\textwidth]{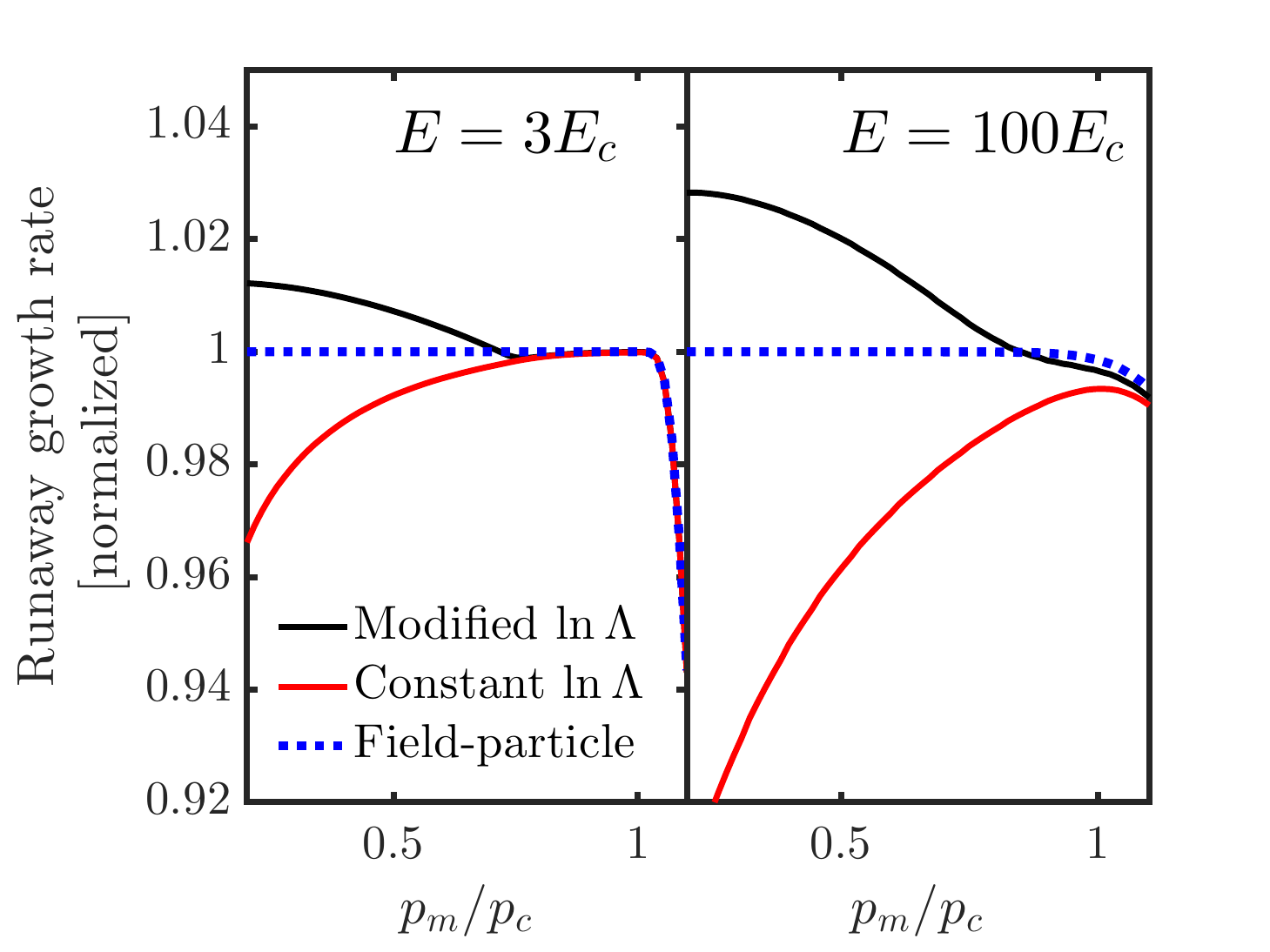}
\caption{\label{fig:pm scan} Runaway growth rate as function of
  momentum cut-off parameter $p\sub{m}$ for two different electric
  fields, normalized to the field-particle $p_m=0$
    value. Lines correspond to (dotted blue) the field-particle
  Boltzmann operator, Eq.~(\ref{eq:full field-particle}), and (solid)
  the full operator including the test-particle operator when (red)
  $\ln\Lambda$ is held fixed or (black) modified according to
  Eq.~(\ref{eq:lnLambda mod}), which is the physically most correct
  model.  Plasma parameters: thermal electron density
  $n\sub{e} = 10^{20}\,$m$^{-3}$; temperature $T\sub{e} = 100\,$eV. }
\end{center}
\end{figure}

We will now demonstrate that our complete knock-on model satisfies the
essential property that the solutions to the kinetic equation are
independent of the arbitrary cut-off momentum $p_m$, as long as it is
chosen small enough.  To determine the sensitivity of the solutions to
$p\sub{m}$, we will consider the instantaneous runaway growth rate
when the primary runaway population is described by a shifted
Maxwellian runaway distribution
$f\sub{RE} \propto \exp\big[-(\boldsymbol{p}-\boldsymbol{p}_0)^2/q^2\big]$. For this
test we have chosen the momentum $p_0 \approx 6 \,m\sub{e}c$ in the
parallel direction, with width $q \approx 0.6 m\sub{e} c$.  Two
electric-field strengths are investigated, a low-field case where
$E=3E\sub{c}$ and a high-field case $E=100 E\sub{c}$.  The resulting
growth rates are shown in Fig.~\ref{fig:pm scan}, as a function of the
cut-off $p\sub{m}$ after a short time $0.03\tau\sub{c}$.
The growth rate obtained using the field-particle operator alone is
nearly independent of $p\sub{m}$ as long as it is smaller than $p_c$,
indicating that secondary particles created with momentum $p<p_c$ are
unlikely to run away. For the Rosenbluth-Putvinski
operator, this behavior was also observed by~\citet{Nilsson2015}.  

When the test-particle operator is added, but the Coulomb logarithm
$\ln\Lambda$ is left unmodified, the growth rate is decreased. This
can be understood from the fact that the test-particle operator
represents a source of energy loss for the runaways, which diverges
logarithmically as $p\sub{m} \to 0$.  When $\ln\Lambda$ is modified
(black line in Fig.~\ref{fig:pm scan}, representing the most
physically accurate model), the mean energy-loss rate of a runaway
becomes independent of $p\sub{m}$. The growth rate, however, is found
to increase with decreasing $p\sub{m}$, settling to a constant value
in the limit $p\sub{m} \to 0$. The underlying mechanism for this
behavior is that a fraction of all collisions are now accounted for
with a Boltzmann operator rather than with a Fokker-Planck
operator. This leads to an increase in the runaway probability for
particles with $p<p_c$, since the Boltzmann operator fully captures
the stochastic nature of the collisions; instead of continuously
experiencing the average energy loss, an electron is accelerated
freely until it undergoes a collision, by which point it may have
gained enough energy to enter the runaway region ($p>p_c$).  Note that
this effect only appears to modify the growth rate with a few percent,
the effect being weaker for smaller electric fields.  The effect is,
however, directly proportional to $1/\ln\Lambda$, as it depends on the
relative importance of small-angle and large-angle collisions. This
implies that for higher-density or lower-temperature plasmas, the
effect can be expected to be more pronounced.

It should be remarked that the field-particle knock-on operator uses a
constant Coulomb logarithm in the Fokker-Planck operator, yet is still
well behaved when $p_m$ becomes small.  We have pointed out that the
field-particle knock-on operators, like those used in previous
runaway-avalanche studies, double count collisions with the
Fokker-Planck operator. However, they do so only with the
\emph{field-particle} Fokker-Planck operator, and not the
test-particle operator which describes the friction on
runaways. Therefore, only the Coulomb logarithm in the field-particle
operator should be modified when using such models.  The
field-particle Fokker-Planck operator is essential when considering
the dynamics of the bulk population, however it does not significantly
affect the avalanche growth rate, thereby explaining the insensitivity
to $p_m$ for $p_m\lesssim p_c$.

\subsection{Steady-state avalanche growth rate at moderate electric fields}
\begin{figure}
\begin{center}
\includegraphics[width=0.6\textwidth]{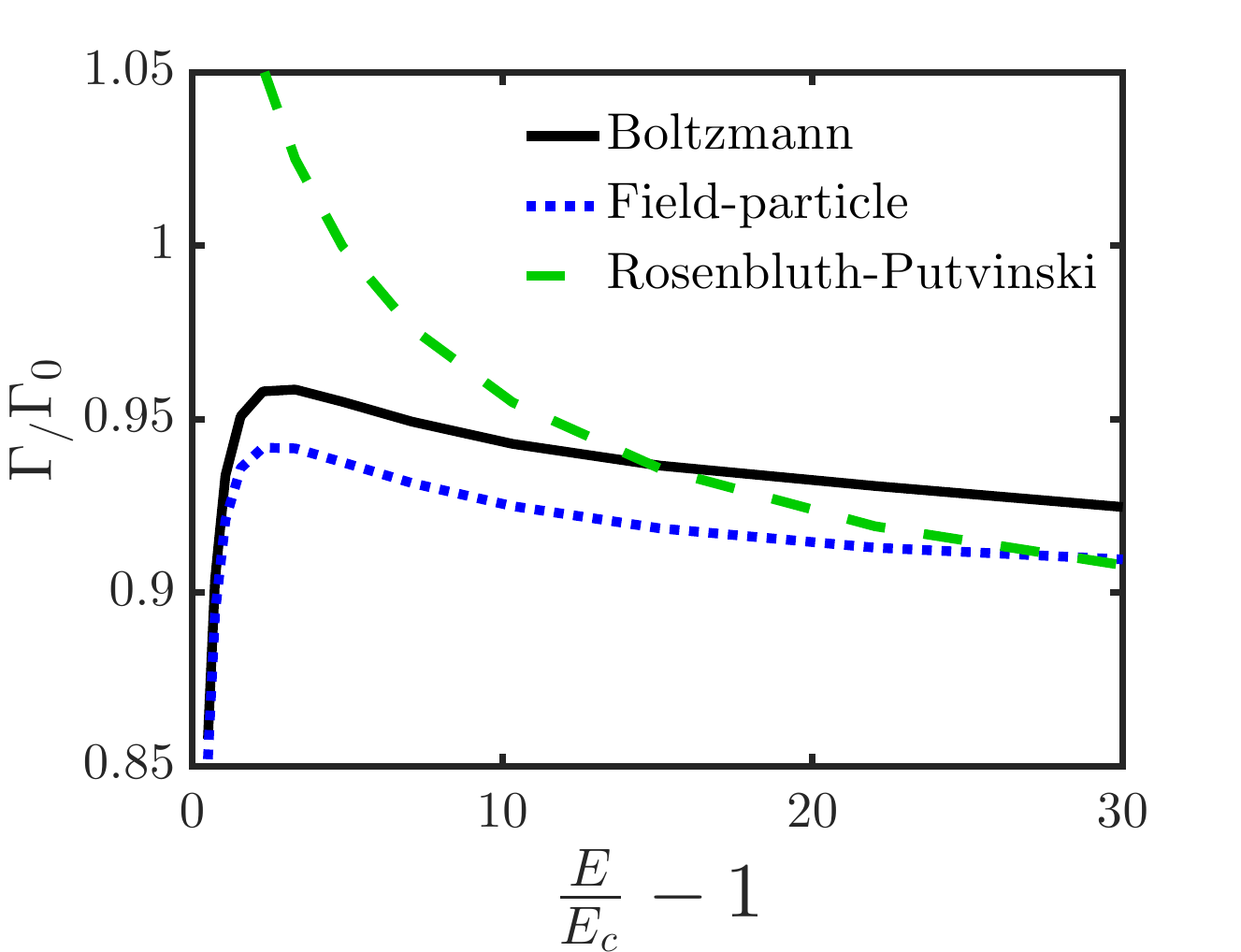}
\caption{\label{fig:steady state growth} Steady-state runaway growth
  rate normalized to the diffusion-free result
  $\Gamma_0 = (E/E\sub{c}-1)/2\ln\Lambda$,
  Eq.~(\ref{eq:ava_growth_rate}), in the presence of a constant
  electric field, neglecting radiation losses.  Plasma parameters:
  thermal electron density $n\sub{e} = 10^{20}\,$m$^{-3}$; temperature
  $T\sub{e} = 100\,$eV; effective charge $Z\sub{eff} = 1$.}
\end{center}
\end{figure}

The steady state avalanche growth rate in a constant electric field is
a classical result; Rosenbluth and Putvinski derived the
growth-rate formula \eqref{eq:ava_growth_rate} in 1997. 
After an initial transient, the distribution function tends
to approach the asymptotic quasi-steady state behavior $f(t,\,p,\,\xi) \sim n\sub{RE}(t)\bar{f}(p,\,\xi)$,
where $\int \bar f \,\rd \boldsymbol{p} = 1$. The kinetic equation, being linear in the runaway distribution, then prescribes that
the runaway population will grow with a constant growth rate 
\begin{align}
\Gamma = \frac{1}{n\sub{RE}}\frac{\rd n\sub{RE}}{\rd (t/\tau_c)}.
\label{eq:gamma}
\end{align}%
In Figure~\ref{fig:steady state growth} we show the growth rate
$\Gamma$ obtained from numerical solutions of the kinetic equation
using various models for the knock-on operator, for moderate
electric fields ranging from $E=1.5E\sub{c}$ to $E=30E\sub{c}$ and $Z\sub{eff} =
1$. We see that using the Rosenbluth-Putvinski knock-on operator leads
to a significant error compared to the more accurate models when the
electric field is near the critical -- of order $30\%$ at $1.5E\sub{c}$. At
larger electric fields the error is insignificant.  Interestingly, the
full Boltzmann operator (solid black line) yields a correction of only a few percent
compared to the field-particle operator alone. This means that
the test-particle part of the operator does not influence the growth
rate significantly. This result is robust; it is not affected by
changes in thermal electron density and temperature, and only slightly
modified by changes in the effective charge. Note that a significant
error is obtained if one fails to account for the double counting of
small-angle collisions -- the size of this error is sensitive to the
cut-off $p\sub{m}$, diverging logarithmically as it approaches zero,
and the result is included primarily for illustrative
purposes.

\subsection{Avalanche generation in a near-threshold electric field with synchrotron-radiation losses}
In tokamaks, the runaway dynamics in electric fields near the runaway-generation threshold is of particular interest. Due to the large self-inductance of tokamaks, after a transient phase during which the ohmic current of the background is dissipated, the electric field will tend towards that value $E_a$ -- the threshold field -- for which the runaway growth rate vanishes, $\Gamma(E_a) = 0$~\citep{Breizman2014}.

At these low electric fields, radiation losses have a large impact, and can not be ignored in the calculation of the runaway growth rate. In this section, we will include the effect of synchrotron-radiation losses and investigate runaway generation when $E\sim E_a$. 
A model for this was recently presented by~\citet{Aleynikov2015} (referred to as A\&B), using a simplified kinetic equation following a method used by~\citet{Lehtinen1999}. An interesting prediction by the A\&B model was that reverse knock-on can have a significant effect on the growth rate, where for electric fields $E\lesssim E_a$, existing runaways will be slowed down to $v<v_c$ in single large-angle collision events. This leads to a negative avalanche-growth rate, which previous large-angle collision models are incapable of describing, as this process is inherently a large-angle \emph{test-particle} effect. Using the knock-on operator presented in this work, we will now assess the magnitude of the reverse-knock on effect, as well as determine the threshold field $E_a$ and the growth rate when $E\sim E_a$, accounting for radiation losses.

\begin{figure*}
\begin{center}
\includegraphics[width=\textwidth, trim=0mm 0 0mm 0mm]{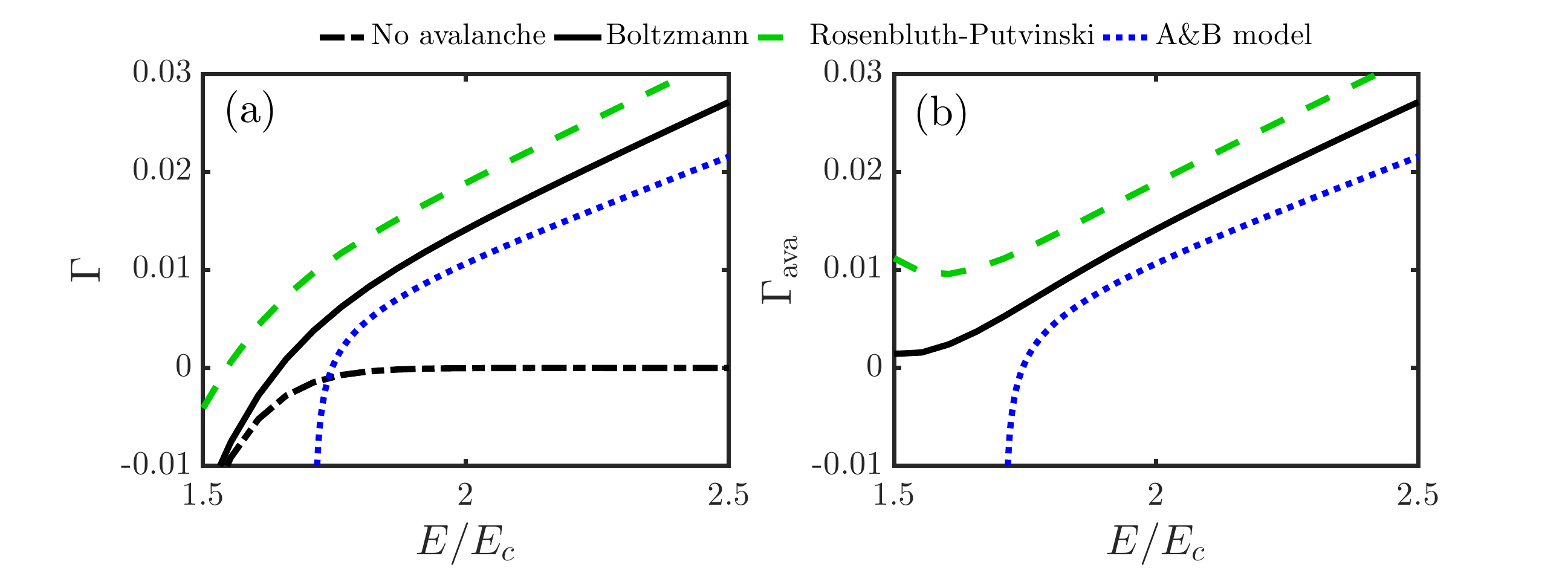}
\caption{\label{fig:synchrotron avalanche growth} Steady-state runaway
  growth rate in the presence of a constant electric field, accounting
  for synchrotron radiation losses and using various models for the
  large-angle collision operator: the Rosenbluth-Putvinski operator,
  Eq.~(\ref{eq:RP source}) (green, dashed); the Boltzmann
  operator Eqs.~(\ref{eq:full test-particle})-(\ref{eq:full
    field-particle}) (black, solid), and without any large-angle collision
  operator (black, dash-dotted). For comparison we have included Eq.~(11)
  of \citet{Aleynikov2015} (blue, dotted). In (b), the avalanche-free
  growth rate (black, dotted line in (a)) has been subtracted to yield
  a pure ``avalanche growth rate'' $\Gamma\sub{ava}$.  Plasma
  parameters: thermal electron density $n\sub{e} = 10^{20}\,$m$^{-3}$;
  temperature $T\sub{e} = 1\,$keV; effective charge $Z\sub{eff} = 5$, $B =
  1.81\,$T. }
\end{center}
\end{figure*}

In Fig.~\ref{fig:synchrotron avalanche growth} we show how the
quasi-steady state growth rate $\Gamma$ depends on the electric-field
strength, similar to Fig.~4 of A\&B. We use the same plasma parameters
$Z\sub{eff} = 5$ and
$\bar\tau\sub{rad} = 3m\sub{e}n\sub{e} \ln\Lambda / (2\varepsilon_0
B^2) = 70$
(corresponding to $B\approx1.81\,$T at
$n\sub{e} = 10^{20}\,$m$^{-3}$), although a slight discrepancy occurs
due to our $\ln\Lambda = 14.9$ -- consistent with the background
parameters chosen -- compared to their $\ln\Lambda=18$. In this
scenario, the A\&B threshold electric field is $E_a \approx 1.71E_c$.
Several models for the knock-on operator are included in the
comparison, in addition to the no-avalanche case since we are now
interested in the sub-threshold dynamics.

\begin{figure}
\begin{center}
\includegraphics[width=0.7\textwidth, trim=0mm 20mm 0mm 0mm]{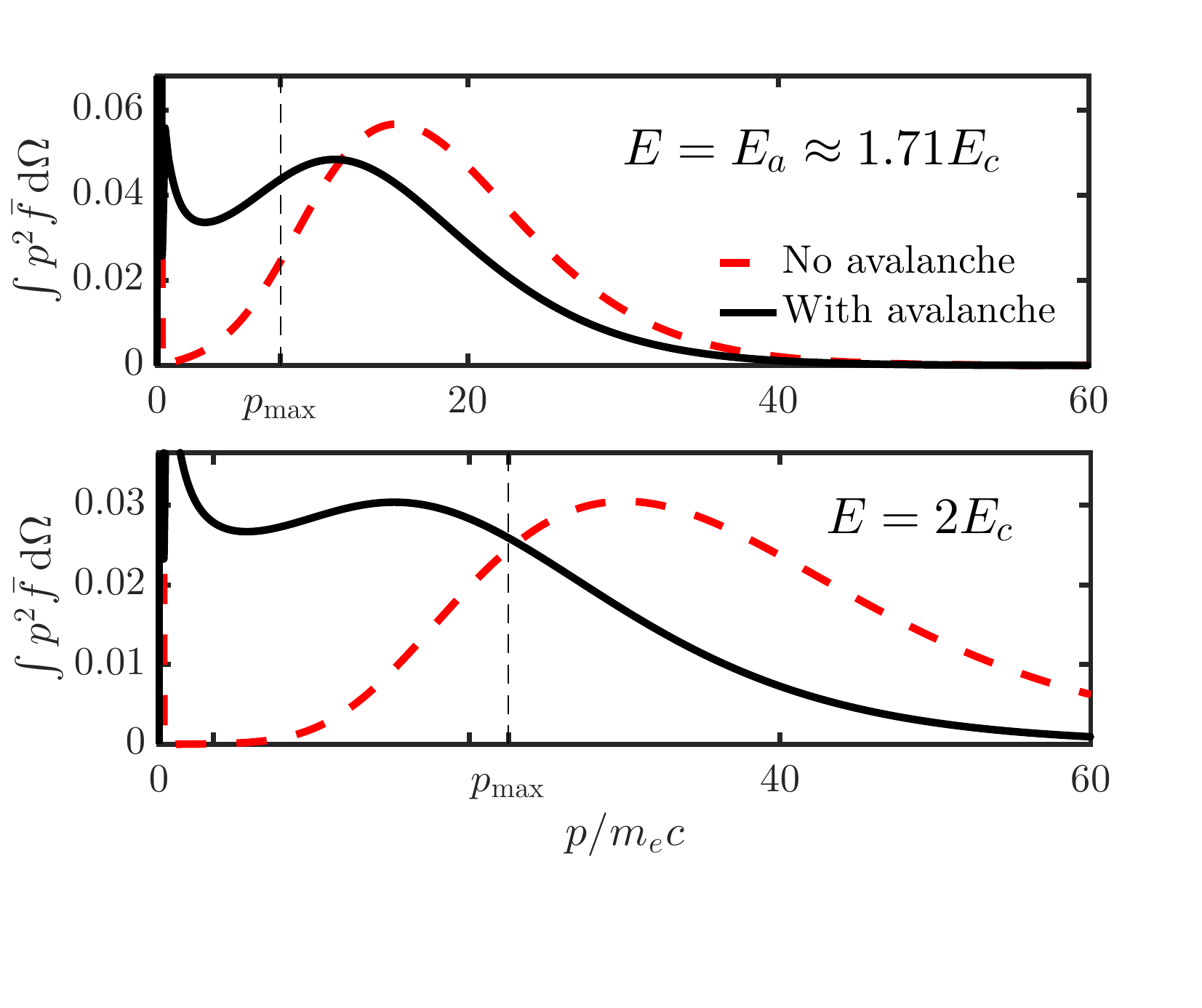}
\caption{\label{fig:synchrotron avalanche fBar} Steady-state runaway
  momentum distributions $\int p^2 \bar{f}\rd\Omega$ (defined to have
  unit area under the shown curves). Included in the figures is the
  maximum runaway momentum $p\sub{max}$ predicted
  by~\citet{Aleynikov2015}.  Plasma parameters: background electron
  density $n\sub{e} = 10^{20}\,$m$^{-3}$; temperature $T\sub{e} = 1\,$keV;
  effective charge $Z\sub{eff} = 5$, $B = 1.81\,$T.}
\end{center}
\end{figure}

It is interesting to observe that the test-particle operator, which
allows runaways to be thermalized in a single large-angle collision,
does not significantly modify the dynamics, in contrast to the
theoretical prediction by~\citep{Aleynikov2015}.  Unlike the A\&B
model, which predicts a significant negative growth rate due to this
effect when $E\lesssim E_a$, we find that the large-angle collision
operator always adds a positive contribution to the total growth rate
compared to the no-avalanche case (see Fig.~\ref{fig:synchrotron
  avalanche growth}(b) where the no-avalanche growth rate has been
subtracted).  It can be concluded that the negative growth rates in
the sub-threshold regime is a result primarily of the Fokker-Planck
dynamics, rather than of large-angle collisions.  The reason for this
discrepancy to the A\&B model can be understood by considering the
behavior of the distribution shape functions $\bar{f}=f/n$, defined
before Eq.~(\ref{eq:gamma}), which are illustrated in
Fig.~\ref{fig:synchrotron avalanche fBar}. A\&B predicted the
distribution to be a delta function in momentum, located at the point
of force balance, $p\sub{max}$.  When this occurs near the critical
speed $v_c$, runaways cannot produce knock-ons with sufficient energy
to become runaway. Large-angle collisions then act only to slow down
the existing population. In numerical solutions of the full kinetic
equation, conversely, it is found that the runaway population takes on
a wide energy spectrum, and there will always be sufficiently many
runaways with the energy required to produce new runaways to counter
the reverse knock-on effect.

A notable difference between the A\&B model and full solutions of the kinetic equation considered here, which can play an
important role when considering the decay of the runaway current in 
tokamaks, is that the growth rate is not as sensitive to variations in 
electric field close to (but below) the effective critical field $E_a$ as
predicted by A\&B. It is known that the decay rate is determined primarily by the value of the
effective critical field $E_a$ when the self-inductance can be
considered large (roughly when the total runaway current is much
larger than $\sim 200\,$kA)~\citep{Breizman2014}.
Since the threshold field $E_a$ given by the A\&B model is reasonably accurate in many cases,
it is likely that it may be used to describe runaway current decay in large-current scenarios.
However, for moderate runaway currents in the range of hundreds of kA, the overall
shape of $\Gamma(E)$ will determine its evolution,
which previous theoretical models fail to describe -- particularly for electric fields $E\lesssim E_a$. 

\begin{figure}
\begin{center}
\includegraphics[width=0.7\textwidth, trim=0mm 0 0mm 0mm]{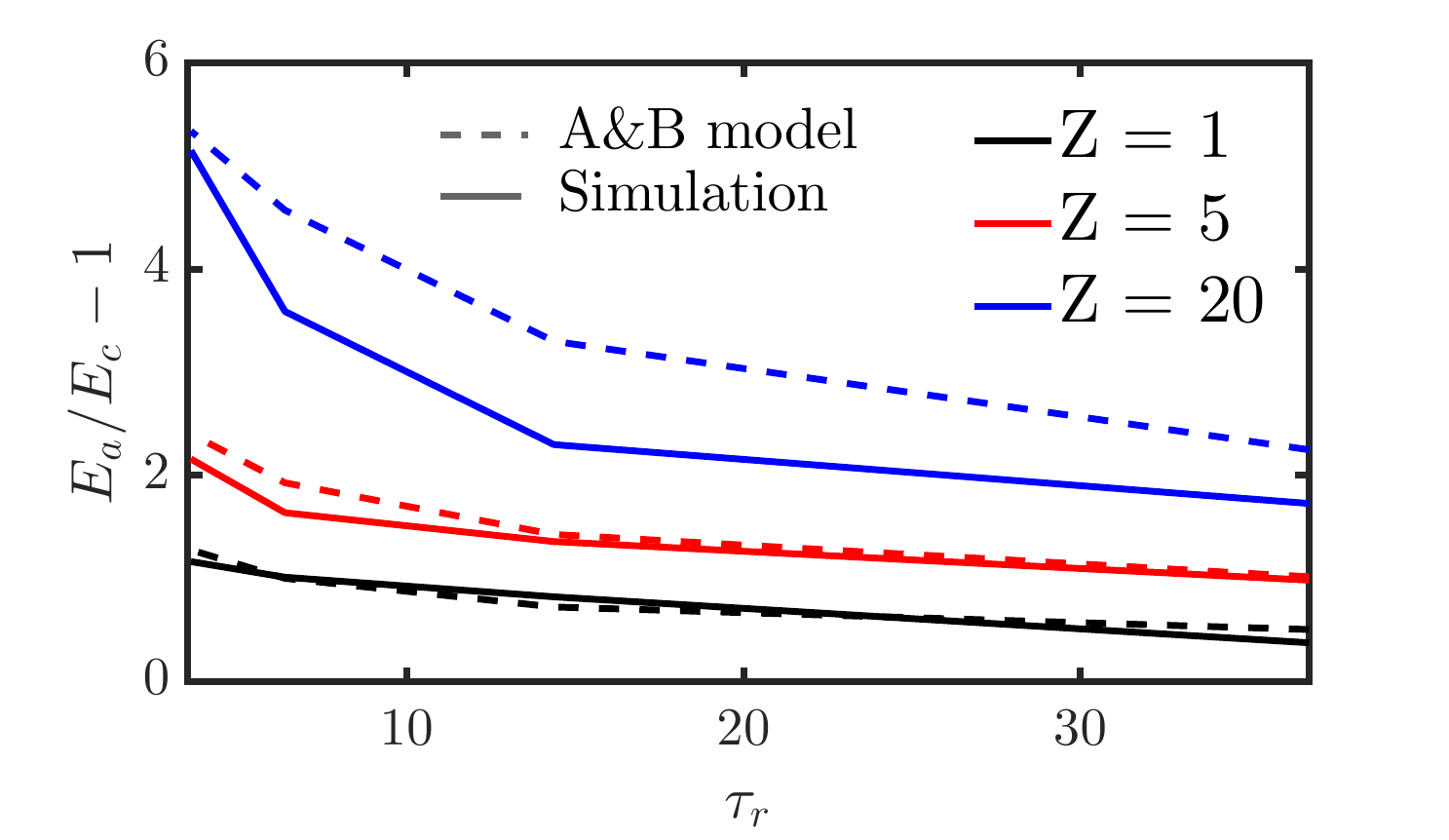}
\caption{\label{fig:Ea scan} Threshold electric field determined
  numerically from solutions of the kinetic equation, as a function of
  normalized magnetic field strength $\tau\sub{r}$ for various values
  of the effective charge. Predictions by the theoretical model
  of~\citet{Aleynikov2015} are included for comparison.}
\end{center}
\end{figure}

Finally, we show the effective critical field $E_a$
calculated numerically by CODE for a wide range of 
$Z\sub{eff}$ and magnetic-field strength parameters
$\tau\sub{r}=6 \upi \epsilon_0^2 m_e^2c^3/e^4B^2\tau_c$. This is shown in Fig.~\ref{fig:Ea scan}, along with
the values given by the A\&B model by determining the
roots of their Eq.~(11). It is seen that the predictions
of \citet{Aleynikov2015} are typically accurate unless the effective
charge is very large, and are most accurate for sufficiently small or
large $B$. The observed trend in the accuracy of their model is
unexpected, since they have utilized fast pitch-angle equilibration
time (large $Z\sub{eff}$) and weak magnetic field (large $\tau\sub{r}$)
in order to reduce the kinetic equation to a tractable form.

\section{Conclusions}

Predictions indicate that a major part of the initial plasma current
in large tokamaks can be converted to runaway electron current. This
is partly due to the large plasma size limiting the loss of
runaway-electron seeds, but more importantly, it is due to the
avalanche mechanism which leads to an exponential growth of
runaways. The runaway-electron growth rate due to avalanching is
exponentially sensitive to the plasma current, and avalanche runaway
generation is therefore expected to be a serious issue in ITER and
other high-current reactor-scale tokamaks. As the plasma current in
present devices cannot be increased above a few megaamperes, full
experimental simulation of high-current tokamak disruptions is not
possible. Therefore it is very important to develop accurate
theoretical models from first principles, to test the validity of
approximative models.

In this paper we have developed a fully conservative knock-on
collision operator derived from the relativistic Boltzmann operator,
and compared it to existing models. Close to the critical electric
field, the new model leads to behavior significantly different from
that of the widely used \citet{RosenbluthPutvinski1997} avalanche
model. This influences the predictions for the transformation of a
runaway seed to an avalanching population; fortunately the new
operator predicts a lower growth rate
than~\citet{RosenbluthPutvinski1997} and therefore the implications
for ITER should be positive, although the difference between models is
marginal for high electric fields.  We have also described how to
resolve the issue of double-counting the small-angle and large-angle
collisions, and have illustrated the importance of this issue.
The new operator includes both the test-particle and field-particle
parts of the collision operator, however we have shown that
the test-particle part does not influence the growth rate
significantly.

Using kinetic simulations we have performed a careful study of the
runaway growth rate in the presence of synchrotron radiation losses
and several different avalanche operators. Again, we find a
significant difference in runaway rates close to the critical field,
however, the effective critical field appears to be well reproduced by
simplified models unless the effective charge is very large.

\begin{acknowledgments}
  The authors would like to thank Linnea Hesslow, Istv\'an Pusztai,
  Eero Hirvijoki, Jack Connor, Gergely Papp and Matt Landreman for
  constructive discussions. This work was supported by the Swedish
  Research Council (Dnr. 2014-5510) and the European Research Council
  (ERC-2014-CoG grant 647121). The project has also received funding
  from the European Union's Horizon 2020 research and innovation
  programme under grant agreement number 633053. The views and
  opinions expressed herein do not necessarily reflect those of the
  European Commission.
\end{acknowledgments}

\appendix
\section{Density, momentum and energy conservation}
\label{ap:conservation}
The full electron-electron Boltzmann operator $C$ is known to satisfy conservation of density, momentum and energy, expressed by the relations
\begin{align}
\int \rd \boldsymbol{p}\, C(f\sub{e}) &= 0, \nonumber \\
\int \rd \boldsymbol{p}\, \boldsymbol{p} C(f\sub{e}) &= 0, \\
\int \rd \boldsymbol{p}\, m\sub{e} c^2 (\gamma-1) C(f\sub{e}) &= 0, \nonumber
\end{align}
or in our case of a cylindrically symmetric plasma, in terms of the Legendre modes of the collision operator,
\begin{align}
\int \rd p \, p^2 C_0(p) &= 0, \nonumber \\
\int \rd p \, p^3 C_1(p) &= 0, \\
\int \rd p \, p^2 (\gamma-1) C_0(p) &= 0. \nonumber 
\end{align}
We will show that our explicit form of the knock-on operator,
accounting only for collisions involving electrons with momenta
$p>p\sub{m}$, satisfy the same conservation laws. Taking the full
operator
$C_L = C_L\{f\sub{e},\,f\sub{Me}\} + C_L\{f\sub{Me},\,f\sub{e}\}$ from
Eqs.~(\ref{eq:full test-particle})-(\ref{eq:full field-particle}),
noting that the gain term only applies for $\gamma > \gamma\sub{m}$
and the loss term for $\gamma > 2\gamma\sub{m}-1$, we find upon
integration (changing momentum integrals to energy integrals by
$v \rd p =m\sub{e} c^2 \rd \gamma$)
\begin{align}
&\Gamma_L\{h\} =  \int_0^\infty \rd p \, p^2 h(p) C_L(p)  \nonumber \\
&= \frac{m\sub{e} c}{2\tau\sub{c} \ln\Lambda} \int_{\gamma\sub{m}}^\infty \rd \gamma \, h(p) \int_{\gamma+\gamma\sub{m}-1}^\infty \rd \gamma_1 \, p_1^2 \Sigma(\gamma,\,\gamma_1)P_L(\xi^*)f_L(p_1) ) \nonumber \\
& - \frac{m\sub{e} c}{4\tau\sub{c} \ln\Lambda} \int_{2\gamma\sub{m}-1}^\infty \rd \gamma \, p^2  h(p) f_L(p) \int_{\gamma\sub{m}}^{\gamma+1-\gamma\sub{m}} \rd \gamma_1 \,\Sigma(\gamma_1,\,\gamma)
\nonumber \\
& - \frac{m\sub{e} c h(0)}{4\tau\sub{c} \ln\Lambda} \delta_{L,0} \int_{2\gamma\sub{m}-1}^\infty \rd \gamma \, p^2 f_0(p) \int_{\gamma\sub{m}}^{\gamma+1-\gamma\sub{m}} \rd \gamma_1 \,\Sigma(\gamma_1,\,\gamma).
\end{align}
In the first term, the integration order can be interchanged by using
\begin{align}
\int_{\gamma\sub{m}}^\infty \rd \gamma  \int_{\gamma+\gamma\sub{m}-1}^\infty \rd \gamma_1 = \int_{2\gamma\sub{m}-1}^\infty \rd \gamma_1 \,\int_{\gamma\sub{m}}^{\gamma_1+1-\gamma\sub{m}}\rd \gamma .
\end{align}
Exchanging the names of the dummy variables $\gamma_1$ and $\gamma$ in
this term then yields
\begin{align}
&\frac{2\tau\sub{c} \ln\Lambda}{m\sub{e} c}\Gamma_L\{h\} = \int_{2\gamma\sub{m}-1}^\infty \rd \gamma \, p^2 f_L(p) \times \nonumber \\
& \times \int_{\gamma\sub{m}}^{\gamma+1-\gamma\sub{m}} \hspace{-3mm} \rd \gamma_1 \, \left[ h(p_1)P_L\left(\frac{\gamma+1}{\gamma_1+1}\frac{p_1}{p}\right) - \frac{h(p) + \delta_{L,0}h(0)}{2}\right]\Sigma(\gamma_1,\,\gamma).
\end{align}
The conservation of density, momentum and energy correspond to the
conditions $0 = \Gamma_0\{1\} = \Gamma_1\{p\} = \Gamma_0\{\gamma-1\}$,
respectively. With $L=0$ and $h = 1$, the bracket term in the
$\gamma_1$-integral vanishes identically; therefore the knock-on
operator will conserve density independently of the differential
cross-section $\Sigma(\gamma_1,\,\gamma) = (2\upi r_0^2)^{-1}\partial
\sigma/\partial \gamma$. For the other two conditions, one finds
\begin{align}
2\tau\sub{c} \ln\Lambda \Gamma_1\{p\} &= (m\sub{e} c)^2 \int_{2\gamma\sub{m}-1}^\infty \rd \gamma \, p(\gamma+1) f_1(p) \nonumber \\
&\hspace{-7mm}\times  \int_{\gamma\sub{m}}^{\gamma+1-\gamma\sub{m}} \rd \gamma_1  \,\left[\gamma_1-1 - \frac{\gamma-1}{2}\right]\Sigma(\gamma_1,\,\gamma), \\
2\tau\sub{c} \ln\Lambda \Gamma_0\{\gamma-1\} &=  m\sub{e} c\int_{2\gamma\sub{m}-1}^\infty \rd \gamma \, p^2 f_0(p) \nonumber \\
&\hspace{-7mm}\times \int_{\gamma\sub{m}}^{\gamma+1-\gamma\sub{m}} \rd \gamma_1  \,\left[\gamma_1-1 - \frac{\gamma-1}{2}\right]\Sigma(\gamma_1,\,\gamma).
\end{align}
The integrals over $\gamma_1$ will vanish for all cross-sections that
respect the indistinguishability of the electrons, i.e.~for which
$\Sigma(\gamma_1,\,\gamma) = \Sigma(\gamma_2,\,\gamma)$ where
$\gamma_2 = \gamma + 1 -\gamma_1$. This follows directly from the
observation that
\begin{align}
\left[\gamma_1-1-\frac{\gamma-1}{2}\right] = -\left[\gamma_2-1-\frac{\gamma-1}{2}\right],
\end{align}
confirming that our operator indeed satisfies the conservation laws.

\subsection{Total cross-section}
For our case of the M\o{}ller cross-section, the differential
cross-section of Eq.~\ref{eq:moller} can be integrated
analytically to produce the total
cross-section.  One obtains
\begin{align}
\sigma(p) &= \int_{\gamma\sub{m}}^{\gamma+1-\gamma\sub{m}} \rd \gamma_1 \, 2\upi r_0^2\Sigma(\gamma_1,\,\gamma) \nonumber \\
& = \frac{2\upi r_0^2}{\gamma^2-1}\Biggr[ \left(\frac{\gamma+1}{2}-\gamma\sub{m}\right)\left(1+\frac{2\gamma^2}{(\gamma-\gamma\sub{m})(\gamma\sub{m}-1)}\right)  \nonumber \\
&\hspace{39mm} - \frac{2\gamma-1}{\gamma-1}\ln\frac{\gamma-\gamma\sub{m}}{\gamma\sub{m}-1} \Biggr].
\label{eq:total sigma}
\end{align}

\section{The Chiu-Harvey and Rosenbluth-Putvinski models}
\label{ap:rpch}

We derive the Chiu-Harvey source by assuming runaways to have a
negligible perpendicular velocity component, i.e.~that $f\sub{e}$ is well
described by a delta function in pitch-angle, $f\sub{e}(p_1,\,\cos\theta_1)
= F(p_1)\delta(\cos\theta_1-1)/(2\upi p_1^2)$ with $F(p_1) = 2\upi
\int_{-1}^1 \rd \cos\theta_1 \, p_1^2 f\sub{e}(p_1,\,\cos\theta_1)$.  From
Eq.~(\ref{eq:generalized S}) we then find
\begin{align}
S\sub{CH} &= \frac{1}{4\upi\tau\sub{c} \ln\Lambda}\frac{1}{p\gamma}\int_{q^*}^\infty \rd p_1 \, \frac{p_1}{\gamma_1}F(p_1)\Sigma(\gamma,\,\gamma_1) \int_{-1}^1 \rd \xi_1\,\delta(\xi_s-\xi^*)\delta(\xi_1-1) \nonumber \\
& =  \frac{1}{4\upi\tau\sub{c} \ln\Lambda}\frac{1}{p\gamma} \int_{q^*}^\infty \rd p_1 \, \frac{p_1}{\gamma_1}F(p_1) \Sigma(\gamma,\,\gamma_1)\delta(\xi- \xi^*) \nonumber \\
&= \frac{1}{4\upi\tau\sub{c} \ln\Lambda}\frac{p_1^2}{p\gamma \xi} F(p_1)\Sigma(\gamma,\,\gamma_1)H(p_1-q^*),
\label{eq:re CH}
\end{align}
where $H(x)$ denotes the Heaviside step function, and we used 
\begin{align}
\left.\frac{\rd p_1}{\rd \xi^*}\right|_{\xi^* = \xi} = \frac{\gamma_1 p_1}{\xi}.
\end{align}
We also  utilized $\cos\theta\sub{s} = \cos\theta$ when $\cos\theta_1 = 1$,
and kinematics constrain the incident momentum $p_1$ according to
Eq.~(\ref{eq:chiu harvey kinematics}).  This result agrees exactly
with the Chiu-Harvey source $S\sub{CH}$ of Eq.~(\ref{eq:CH source}).
In terms of an expansion in Legendre polynomials, the Chiu-Harvey
avalanche source is obtained from the general field-particle operator
in Eq.~(\ref{eq:full field-particle}) simply by replacing $f_L(p)$ by
$(2L+1)f_0(p)$, corresponding to the delta-function approximation. In
this representation, however, the approximation holds limited appeal
as it does not provide a significant simplification of the collision
operator; indeed, compared to the full operator it requires a larger
number of Legendre polynomials to be retained since the true $f_L$
decreases rapidly with $L$ for sufficiently large $L$.

By the addition of the sink terms in (\ref{eq:full test-particle}) and
(\ref{eq:full field-particle}), and extending the integration limit
down from $q^*$ to $q_0$, the Chiu-Harvey operator can be made
conservative.  However, the delta-function assumption in pitch-angle
causes incorrect momentum dynamics, and the total momentum of the
distribution will not be conserved in this treatment.  This can be
corrected by treating the $L=1$ mode exactly, corresponding to a total
conservative knock-on operator in the Chiu-Harvey approximation of the
form
\begin{align}
C\sub{CH}^{\text{(cons)}} &= \bar{S}\sub{CH} - \frac{1}{4\tau\sub{c} \ln\Lambda} v f\sub{e}(\boldsymbol{p})\sigma(p) - \frac{\delta(\boldsymbol{p})}{4\tau\sub{c} \ln\Lambda} \int_{p' > q_0(p\sub{m})}\hspace{-8mm} \rd \boldsymbol{p}'\, v'f\sub{e}(\boldsymbol{p}') \sigma(p') \nonumber \\
& - \frac{3(m\sub{e} c)^{-3}}{8\upi\tau\sub{c} \ln\Lambda}\frac{\xi}{\gamma (\gamma+1)}\int_{p_1>q_0} \hspace{-3mm} \rd \boldsymbol{p}_1 \, \frac{\gamma_1+1}{\gamma_1} \Sigma(\gamma,\,\gamma_1)
(1-\xi_1) f\sub{e}(\boldsymbol{p}_1) \nonumber \\
&+ \frac{3}{8\tau\sub{c} \ln\Lambda} \xi v \sigma(p)\int_{-1}^1 \rd \xi_1 \, (1-\xi_1)f\sub{e}(p,\,\xi_1),
\end{align}
where $\bar{S}\sub{CH}$ equals (\ref{eq:re CH}) with $q^*$ changed to
$q_0$, and the last two momentum-correcting terms are small when the
runaway population consists predominantly of electrons with small
pitch-angle, $1-\xi_1 \ll 1$. Unlike the Chiu-Harvey model, this
operator depends not only on the angle averaged distribution $\int f\sub{e}
\, \rd \xi$, but also on $\int(1-\xi)f\sub{e}\,\rd\xi$.

Note that the issue of double-counting is important only in the
test-particle part of the operator. In the Chiu-Harvey approach, when
the test-particle part is neglected, only field-particle collisions
would be double-counted, and for those the small-angle collisions have
negligible impact on runaway generation.

The Rosenbluth-Putvinski result is obtained under the assumptions that
the primary electrons not only have small pitch-angle, but also large
energy. Therefore, in the second line of (\ref{eq:re CH}),
$(p_1/\gamma_1)\Sigma(\gamma,\,\gamma_1)\delta(\xi-\xi^*)$ can be
replaced by $m\sub{e} c\Sigma(\gamma,\,\infty)\delta(\xi-\xi_0)$ with an
error of order $1/\gamma_1$, where $\xi_0 = \lim_{\gamma_1\to\infty}
\xi^* = \sqrt{(\gamma-1)/(\gamma+1)}$. Under this assumption, the
source term reduces to
\begin{align}
S\sub{RP} &= \frac{m\sub{e} c}{4\upi\tau\sub{c}\ln\Lambda}\frac{\delta(\xi-\xi_0)}{p\gamma}\Sigma(\gamma,\,\infty)\int_{q^*}^\infty \rd p_1\,F(p_1) + \Ordo(1/\gamma_1) \\
&\approx  \frac{n\sub{RE}}{4\upi\tau\sub{c}\ln\Lambda}\delta(\xi-\xi_0)\frac{m\sub{e}^3 c^3}{p^2}\frac{\rd}{\rd p}\frac{1}{1-\gamma},
\end{align}
where the last step follows by replacing the integral over $F$ by the
runaway number density $n\sub{RE}$, valid when $\gamma^* =
\sqrt{(q^*/m\sub{e} c)^2+1} = 2\gamma-1$ is small compared to typical
runaway energies. This result agrees exactly with $S\sub{RP}$ in
Eq.~(\ref{eq:RP source}). Note that this final approximation allows
secondary electrons to be created with momentum and energy larger than
that of any present primary electron. In fact, when integrated over
all momenta the Rosenbluth-Putvinski source term is found to create
energy and momentum at an infinite rate.

\bibliographystyle{jpp}
\bibliography{closecoll3}

\end{document}